\documentclass{ws-p9-75x6-50}

\def\gsim{\mathrel{%
\rlap{\raise 0.511ex \hbox{$>$}}{\lower 0.511ex
\hbox{$\sim$}}}}

\def\lsim{\mathrel{
\rlap{\raise 0.511ex \hbox{$<$}}{\lower 0.511ex
\hbox{$\sim$}}}}

\begin{document}

\title{Quantum strings and black holes}
\author{Thibault Damour}
\address{Institut des Hautes Etudes Scientifiques, 91440
Bures-sur-Yvette, France\\E-mail: damour@ihes.fr}

\maketitle

\abstracts{
The transition between (non supersymmetric) quantum string states and
Schwarzschild black holes is discussed. This transition occurs when the
string coupling $g^2$ (which determines Newton's constant) increases 
beyond a certain critical value $g_c^2$. We review a calculation   
showing that self-gravity causes a typical string 
state of mass $M$ to shrink, as the string coupling $g^2$ increases, down to 
a compact string state whose mass, size, entropy and luminosity match (for 
the critical value $g_c^2 \sim (M \, \sqrt{\alpha'})^{-1}$) those of a 
Schwarzschild black hole. This confirms the idea (proposed by several
authors) that the entropy of black 
holes can be accounted for by counting string states.
The level spacing of the quantum states of Schwarzschild black holes is 
expected to be exponentially smaller than their radiative width.
This makes it very difficult to conceive (even Gedanken) experiments probing
the discreteness of the quantum energy levels of black holes.}


\section{Introduction}
String theory is the only known theory which reconciles perturbative gravity
with quantum mechanics. It is important to understand the consequences of
string theory for non-perturbative aspects of gravity such as black holes
and cosmological singularities. The basic idea discussed here is that there
is a transition (similar to a phase transition) between certain quantum
states of string theory and  (to be defined) quantum black holes states.
This transition occurs when the string coupling constant $g^2$ 
(which determines Newton's constant, $ G \propto g^2 $) increases 
beyond a certain critical value $g_c^2$.

 Most of the stringy literature 
has concentrated (for reasons recalled below) on some special, 
supersymmetric extreme black holes (BPS black holes). These black holes  
carry special (Ramond-Ramond) charges and their microscopic structure seem 
to be describable (when  $g^2 < g_c^2$) in terms of Dirichlet-branes.
 By contrast, we consider here the 
simplest, Schwarzschild black holes (in any space dimension $d$). It will be 
argued that their ``microscopic structure'' at low $g^2$
 involves only fundamental string 
states. However, the lack of supersymmetry means that it becomes essential 
to deal with self-gravity effects.

To start with, let us recall that thirty years ago the study 
of the spectrum of string theory 
revealed~\cite{FVBM} a huge degeneracy of states growing as an 
exponential of the mass. A few years later Bekenstein~\cite{Bekenstein} 
proposed that the entropy of a black hole should be proportional to the area 
of its horizon in Planck units, and Hawking~\cite{Hawking} fixed the 
constant of proportionality after discovering that black holes do emit 
thermal 
radiation at a temperature $T_{\rm Haw} \sim R_{\rm BH}^{-1}$.

When string and black hole entropies are compared one immediately
notices a striking difference: string entropy\footnote{As we shall discuss, 
the self-interaction of a string lifts the huge degeneracy of free string 
states. One then defines the entropy of a narrow band of string 
states, defined with some energy resolution $M_s \lsim \Delta \, E 
\ll M$, as the logarithm of the number of states within the band 
$\Delta \, E$.} is proportional to the first power of mass in any 
number of spatial dimensions $d$, while black hole entropy is 
proportional to a $d$-dependent power of the mass, always larger 
than $1$. In formulae:
\begin{equation}
S_s \sim {\alpha' M \over \ell_s} \sim M / M_s ~~~~ , ~~~~~~
 S_{\rm BH} \sim \frac{{\rm Area}}{G_N} \sim \frac{R_{\rm
BH}^{d-1}}{G_N} \sim \frac{(g^2 \, M / 
M_s)^{\frac{d-1}{d-2}}}{g^2}\; ,
\label{entropies}
\end{equation}
where, as usual, $\alpha'$ is the inverse of the classical string
tension, $\ell_s \sim \sqrt{\alpha' \hbar}$ is the quantum length 
associated with it\footnote{Below, we shall use the precise 
definition $\ell_s \equiv \sqrt{2 \alpha' \hbar}$, but, in this 
section, we neglect factors of order unity.}, $M_s \sim \sqrt{\hbar 
/ \alpha'}$ is the corresponding string mass scale, $R_{\rm BH}$ is 
the Schwarzschild radius associated with $M$:
\begin{equation}
R_{\rm BH} \sim (G_N \, M)^{1/(d-2)} \; ,\label{eq1.1}
\end{equation}
and we have used that, at least at sufficiently small coupling, the 
Newton constant and $\alpha'$ are related via the string coupling by
$G_N \sim g^2 (\alpha')^{(d-1)/2}$ (more geometrically, 
$\ell_P^{d-1}\sim g^2 \ell_s^{d-1}$).

Given their different mass dependence, it is obvious that, for a 
given set of the  fundamental constants $G_N, \alpha', g^2$,
$S_s > S_{\rm BH}$ at sufficiently small $M$, while the opposite is 
true at sufficiently large $M$. Obviously, there has to be a 
critical value of $M$, $M_c$, at which $S_s = S_{\rm BH}$. This 
observation led Bowick et al.~\cite{Bowick} to conjecture that large 
black holes end up their Hawking-evaporation process when $M = M_c$, 
and then transform into a higher-entropy string state without ever 
reaching the singular zero-mass limit. This reasoning is confirmed  
\cite{GVFC} by the observation that, in string theory, the 
fundamental string length $\ell_s$ should set a minimal value for 
the Schwarzschild radius of any black hole (and thus a maximal value 
for its Hawking temperature). It was also noticed~\cite{Bowick}, 
\cite {susskind},~\cite{GVDivonne} that, precisely at $M= M_c$, 
$R_{\rm BH} = \ell_s$ and the Hawking temperature equals the 
Hagedorn temperature of string theory. For any $d$, the value of 
$M_c$ is given  by:
\begin{equation}
M_c \, \sim M_s g^{-2} \, . \label{eq1.2}
\end{equation}

Susskind and collaborators~\cite 
{susskind},~\cite{halyo} went a step further and proposed that the 
spectrum of black holes and the 
spectrum of single string states be ``identical'', in the sense 
that there be a one to one correspondence between (uncharged) 
fundamental string states and (uncharged) black hole states. Such a 
``correspondence principle'' has been generalized by Horowitz and 
Polchinski~\cite{hp1} to a wide range of charged black hole states 
(in any dimension). Instead of keeping fixed the fundamental 
constants and letting $M$ evolve by evaporation, as considered 
above, one can (equivalently) describe the physics of this 
conjectured correspondence by following a narrow band of states, on 
both sides of and through, the string $\rightleftharpoons$ black 
hole transition, by keeping fixed the entropy\footnote{One uses here 
the fact that, during an adiabatic variation of $g$, the entropy of 
the black hole $S_{\rm BH} \sim ({\rm Area}) / G_N \sim R_{\rm 
BH}^{d-1} / G_N$ stays constant. This result (known to hold in the 
Einstein conformal frame) applies also in string units because 
$S_{\rm BH}$ is dimensionless.} $S = S_s = S_{\rm BH}$, while 
adiabatically\footnote{The variation of $g$ can be seen, depending 
on one's taste, either as a real, adiabatic change of $g$ due to a 
varying dilaton background, or as a mathematical way of following 
energy states.} varying the string coupling $g$, i.e. the ratio 
between $\ell_P$ and $\ell_s$. The correspondence principle then 
means that if one increases $g$ each (quantum) string state should 
turn into a (quantum) black hole state at sufficiently strong 
coupling, while, conversely, if $g$ is decreased, each black hole 
state should ``decollapse'' and transform into a string state at 
sufficiently weak coupling. For all the reasons mentioned above, it 
is very natural to expect that, when starting from a black hole state, 
the critical value of $g$ at which a 
black hole should turn into a string is given, in clear relation to
(\ref{eq1.2}), by
\begin{equation}
g_c^2 \, M \sim M_s \, , \label{eq1.2'}
\end{equation}
and is related to the common value of string and black-hole entropy 
via
\begin{equation}
g_c^2 \sim \frac{1}{S_{\rm BH}} = \frac{1}{S_s}\; . \label{eq1.2''}
\end{equation}
Note that $g_c^2 \ll 1$ for the very massive states ($M \gg
M_s$) that we consider. This justifies our use of the perturbative 
relation between $G_N$ and $\alpha'$.

In the case of extremal BPS, and nearly extremal, black holes
the conjectured correspondence was dramatically confirmed
through the work of  Strominger and Vafa~\cite{SV} and others
\cite{others} leading to a statistical mechanics interpretation of
black-hole entropy in terms of the number of microscopic states 
sharing the same macroscopic quantum numbers. However, little is 
known about whether and how the correspondence works for 
non-extremal, non BPS black holes, such as the simplest  
Schwarzschild black hole\footnote{For simplicity, we shall
consider in this work only Schwarzschild black holes, in any number 
$d \equiv D-1$ of non-compact spatial dimensions.}. By contrast to BPS 
states whose mass is protected by supersymmetry, we shall consider here 
the effect of varying $g$ on the mass and size of non-BPS string states. 

Although it is remarkable that black-hole and string entropy
coincide when $R_{\rm BH} = \ell_s$, this is still not quite
sufficient to claim that, when starting from a string state, a string 
becomes a black hole at $g = g_c$.
In fact, the process in which one starts from a string state in flat 
space and increases $g$ poses a serious puzzle.~\cite{susskind} 
Indeed, the radius of a typical excited string state of mass $M$ is 
generally thought of being of order
\begin{equation}
R_s^{\rm rw} \sim \ell_s (M / M_s)^{1/2} \, , \label{eq1.5}
\end{equation}
as if a highly excited string state were a random walk made of
$M/M_s = \alpha'M/\ell_s$ segments of length $\ell_s$.~\cite{rw} 
[The number of steps in this random walk is, as is natural, the 
string entropy (\ref{entropies}).] The ``random walk'' radius 
(\ref{eq1.5}) is much larger than the Schwarzschild radius for all 
couplings $g \le g_c$, or, equivalently, the ratio of 
self-gravitational binding energy to mass (in $d$ spatial 
dimensions)
\begin{equation}
\frac{G_N \, M}{(R_s^{\rm rw})^{d-2}} \sim \left( \frac{R_{\rm BH}
(M)}{R_s^{\rm rw}} \right)^{d-2} \sim g^2 \left( \frac{M}{M_s}
\right)^{\frac{4-d}{2}} \label{eq1.6}
\end{equation}
remains much smaller than one (when $d>2$, to which we restrict
ourselves) up to, and including, the transition point. In view of 
(\ref{eq1.6}) 
it does not seem natural to expect that a string state will 
``collapse'' to a black hole when $g$ reaches the value 
(\ref{eq1.2'}). One would expect a string state of mass $M$ to turn 
into a black hole only when its typical size is of order of $R_{\rm 
BH} (M)$ (which is of order $\ell_s$ at the expected transition 
point (\ref{eq1.2'})). According to Eq.~(\ref{eq1.6}), this seems to 
happen for a value of $g$ much larger than $g_c$.

Horowitz and Polchinski~\cite{hp2} have addressed this puzzle by
means of a ``thermal scalar'' formalism.~\cite{chi} Their results
suggest a resolution of the puzzle when $d=3$ (four-dimensional
spacetime), but lead to a rather complicated behaviour when $d \geq
4$. Moreover, even in the simple $d=3$ case, the formal nature of the 
auxiliary ``thermal scalar'' renders unclear (at least to me) the physical 
interpretation of their analysis.

Here, I will review the results of a recent collaboration with G.~Veneziano 
\cite{dv} whose aim was to clarify the string 
$\rightleftharpoons$ black hole transition by a direct study, in 
real spacetime, of the size and mass of a {\it typical} excited 
string, within the microcanonical ensemble of {\it self-gravitating} 
strings. Our results~\cite{dv} lead to a rather simple picture of the 
transition, in any dimension. We find no hysteresis
phenomenon in higher dimensions. The critical value for the 
transition is (\ref{eq1.2'}), or (\ref{eq1.2''}) in terms of the 
entropy $S$, for both directions of the string 
$\rightleftharpoons$ black hole transition. In three spatial 
dimensions, we find that the size (computed in real
spacetime) of a {\it typical self-gravitating} string is given by 
the random walk value (\ref{eq1.5}) when $g^2 \le g_0^2$, with 
$g_0^2 \sim (M/M_s)^{-3/2} \sim S^{-3/2}$, and by
\begin{equation}
R_{\rm typ} \sim \frac{1}{g^2 \, M} \, , \label{eq1.8}
\end{equation}
when $g_0^2 \le g^2 \le g_c^2$. Note that $R_{\rm typ}$ smoothly
interpolates between $R_s^{\rm rw}$ and $\ell_s$. This result 
confirms the picture proposed by Ref.~\cite{hp2} when $d=3$, but 
with the bonus that Eq.~(\ref{eq1.8}) refers to a radius which is estimated 
directly 
in physical space (see below), and which is the size of a typical 
member of the microcanonical ensemble of self-gravitating strings. 
In all higher dimensions\footnote{With the proviso that the 
consistency of our analysis is open to doubt when $d\geq 8$.}, we 
find that the size of a typical self-gravitating string remains 
fixed at the random walk value (\ref{eq1.5}) when $g \le
g_c$. However,  when $g$ gets close to a value of order $g_c$, the 
ensemble of self-gravitating strings becomes (smoothly in $d=4$, but 
suddenly in $d \geq 5$) dominated by very compact strings of size 
$\sim \ell_s$ (which are then expected to collapse with a slight 
further increase of $g$ because the dominant size is only slightly 
larger than the Schwarzschild radius at $g_c$).

Our results~\cite{dv} confirm and clarify the main idea of a correspondence
between string states and black hole states~\cite{susskind},
\cite{halyo},~\cite{hp1},~\cite{hp2}, and suggest that the 
transition between these states is rather smooth, with no apparent 
hysteresis, and with continuity in entropy, mass, typical size, and 
luminosity. 
It is, however, beyond the technical grasp of our analysis to 
compute any precise number at the transition (such as the famous 
factor $1/4$ in the Bekenstein-Hawking entropy formula).

\section{Size distribution of free string states}

For simplicity, we 
deal with open bosonic strings ($\ell_s \equiv \sqrt{2 \, \alpha'}$, 
$0 \leq \sigma \leq \pi$)
\begin{equation}
X^{\mu} (\tau , \sigma) = X_{\rm cm}^{\mu} (\tau , \sigma) +
\widetilde{X}^{\mu} (\tau , \sigma) \, , \label{eq2.4}
\end{equation}
\begin{equation}
X_{\rm cm}^{\mu} (\tau , \sigma) = x^{\mu} + 2 \, \alpha' \, p^{\mu} \,
\tau \, , \label{eq2.5}
\end{equation}
\begin{equation}
\widetilde{X}^{\mu} (\tau , \sigma) = i \, \ell_s \sum_{n \not= 0} \
\frac{\alpha_n^{\mu}}{n} \ e^{-i n \tau} \, \cos \, n \, \sigma \, .
\label{eq2.6}
\end{equation}
Here, we have explicitly separated the center of mass motion $X_{\rm
cm}^{\mu}$ (with $[x^{\mu} , p^{\nu}] = i \, \eta^{\mu \nu}$) from 
the oscillatory one $\widetilde{X}^{\mu}$ ($[\alpha_m^{\mu} ,
\alpha_n^{\nu}] = m \, \delta_{m+n}^0 \, \eta^{\mu \nu}$). The free
spectrum is given by $\alpha' \, M^2 = N-1$ where $(\alpha \cdot 
\beta\equiv \eta_{\mu \nu} \, \alpha^{\mu} \, \beta^{\nu} \equiv - 
\alpha^0\, \beta^0 + \alpha^i \, \beta^i)$
\begin{equation}
N = \sum_{n=1}^{\infty} \ \alpha_{-n} \cdot \alpha_n =
\sum_{n=1}^{\infty} \ n \, N_n \, . \label{eq2.7}
\end{equation}
Here $N_n \equiv a_n^{\dagger} \cdot a_n$ is the occupation number 
of the $n^{\rm th}$ oscillator ($\alpha_n^{\mu} = \sqrt{n} \ 
a_n^{\mu}$,$[a_n^{\mu} , a_m^{\nu \dagger}] = \eta^{\mu \nu} \, 
\delta_{nm}$, with $n,m$ positive).

The decomposition (\ref{eq2.4})--(\ref{eq2.6}) holds in any 
conformal gauge ($(\partial_{\tau} \, X^{\mu} \pm \partial_{\sigma} 
\, X^{\mu})^2 = 0$). One can further specify the choice of 
worldsheet coordinates by imposing
\begin{equation}
n_{\mu} \, X^{\mu} (\tau , \sigma) = 2 \alpha' (n_{\mu} \, p^{\mu})
\, \tau \, , \label{eq2.8}
\end{equation}
where $n^{\mu}$ is an arbitrary timelike or null vector ($n \cdot n
\leq 0$).~\cite{scherk} Eq.~(\ref{eq2.8}) means that the 
$n$-projected oscillators $n_{\mu} \, \alpha_m^{\mu}$ are set equal 
to zero. As we shall be interested in quasi-classical, very massive string 
states ($N \gg 1$) it should be possible to work in the ``center of 
mass'' gauge, where the vector $n^{\mu}$ used in Eq.~(\ref{eq2.8}) 
to define the $\tau$-slices of the world-sheet is taken to be the 
total momentum $p^{\mu}$ of the string. This gauge is the most 
intrinsic way to describe a string in the classical limit. Using 
this intrinsic gauge, one can covariantly
define the proper rms size of a massive string state as
\begin{equation}
R^2 \equiv \frac{1}{d} \ \langle (\widetilde{X}_{\perp}^{\mu} \, 
(\tau, \sigma))^2 \rangle_{\sigma , \tau} \, , \label{eq2.9}
\end{equation}
where $\widetilde{X}_{\perp}^{\mu} \equiv \widetilde{X}^{\mu} -
p^{\mu} (p \cdot \widetilde{X}) / (p \cdot p)$ denotes the 
projection of $\widetilde{X}^{\mu} \equiv X^{\mu} - X_{\rm cm}^{\mu} 
(\tau)$ orthogonally to $p^{\mu}$, and where the angular brackets 
denote the (simple) average with respect to $\sigma$ and $\tau$. 

In the center of mass gauge, $p_{\mu} \, \widetilde{X}^{\mu}$ 
vanishes by definition, and Eq.~(\ref{eq2.9}) yields simply
\begin{equation}
R^2 = \frac{1}{d} \, \ell_s^2 \, {\cal R} \, , \label{eq2.10}
\end{equation}
with (after discarding a logarithmically infinite, but state independent, 
contribution)
\begin{equation}
{\cal R} \equiv \sum_{n=1}^{\infty} \ \frac{\alpha_{-n} \cdot 
\alpha_n}{n^2} =  \sum_{n=1}^{\infty} \ \frac{a_n^{\dagger} \cdot 
a_n}{n}= \sum_{n=1}^{\infty} \ \frac{N_n}{n} \, . \label{eq2.11}
\end{equation}
We wish to estimate the distribution function in size of the ensemble of 
free string states of mass $M$, i.e. to count the number of string states, 
having some fixed values of $M$ and $R$ (or, equivalently, $N$ and ${\cal 
R}$). An approximate estimate of this number (``degeneracy'') is~\cite{dv}
\begin{equation}
{\cal D} \, (M,R) \sim \exp \, [ c \, (R) \, a_0 \, M ] \, ,
\label{eq2.30}
\end{equation}
where $a_0 = 2 \, \pi \, ((d-1) \, \alpha' / 6)^{1/2}$ and
\begin{equation}
c \, (R) = \left( 1 - \frac{c_1}{R^2} \right) \left( 1 - c_2 \,
\frac{R^2}{M^2} \right) \, , \label{eq2.31}
\end{equation}
with the coefficients $c_1$ and $c_2$ being of order unity in string
units. The coefficient $c \, (R)$ gives the 
fractional reduction in entropy brought by imposing a size 
constraint. Note that (as expected) this reduction is minimized when 
$c_1 \, R^{-2} \sim c_2 \, R^2 / M^2$, i.e. for $R \sim R_{\rm rw} 
\sim \ell_s \, \sqrt{M/M_s}$.

\section{Mass shift of string states due to self-gravity}

We also need to estimate the mass shift of string states 
(of mass $M$ and size $R$) due to the exchange of the various 
long-range fields which are universally coupled to the string: 
graviton, dilaton and axion. As we are interested in very massive 
string states, $M \gg M_s$, in extended configurations, $R \gg 
\ell_s$, we expect that massless exchange dominates the 
(state-dependent contribution to the) mass shift. [The exchange of spin 1 
fields 
(for open strings) becomes negligible when $M \gg M_s$ because it does not 
increase with $M$.]

The evaluation, in string theory, of (one loop) mass shifts for
massive states is technically quite involved, and can only be 
tackled for the states which are near the leading Regge 
trajectory.~\cite{mshift}
 [Indeed, the vertex operators creating these states 
are the only ones to admit a manageable explicit oscillator
representation.] As we consider states which are very far from the
leading Regge trajectory, there is no hope of computing exactly (at
one loop) their mass shifts. 

In Ref.~\cite{dv} we could estimate the one-loop mass-shift by resorting to 
a semi-classical approximation. The starting point of this semi-classical 
approximation is the effective action of self-gravitating fundamental 
strings derived in Ref.~\cite{BD98}. Using coherent-state methods 
\cite{AABO},~\cite{scherk},~\cite{GSW} and a generalization of Bloch's 
theorem (see Eq.~(3.13) of~\cite{dv}) one finds
\begin{equation}
\delta \, M \simeq - c_d \, G_N \, 
\frac{M^2}{R^{d-2}}\, , \label{eq3.25}
\end{equation}
with the (positive) numerical constant
\begin{equation}
c_d = \left[ \frac{d-2}{2} \, (4\pi)^{\frac{d-2}{2}} \right]^{-1} \, 
,\label{eq3.26}
\end{equation}
equal to $1 / \sqrt{\pi}$ in $d=3$.

The result (\ref{eq3.25}) was expected in order of magnitude, but it is 
important to check that it approximately comes out of a detailed
calculation of the mass shift which incorporates both relativistic 
and quantum effects and which uses the precise definition (\ref{eq2.11}) of 
the squared size.

Finally, let us mention that, by using the same tools Ref.~\cite{dv} has 
computed the imaginary part of the mass shift $\delta \, M = 
\delta \,M_{\rm real} - i \, \Gamma / 2$, i.e. the total decay rate 
$\Gamma$ in massless quanta, as well as the total power radiated $P$. In 
order of magnitude these quantities are
\begin{equation}
\Gamma \sim g^2 \, M \ , \ P \sim g^2 \, M \, M_s \, . 
\label{eq3.29}
\end{equation}

\section{Entropy of self-gravitating strings}
Finally one combines the results of the 
previous sections, Eqs.~(\ref{eq2.30}) and (\ref{eq3.25}), and 
heuristically extend them at the limit of their domain of validity. 
We consider a narrow band of string states that we follow when 
increasing adiabatically the string coupling $g$, starting from $g = 
0$. Let $M_0$, $R_0$ denote the ``bare'' values (i.e. for $g \rightarrow 
0$) of the mass and size of this band of states. Under the adiabatic 
variation of $g$, the mass and size, $M$, $R$, of this band of 
states will vary. However, the entropy$S(M,R)$ remains constant 
under this adiabatic process: $S(M,R) = S (M_0 , R_0)$. We consider states 
with 
sizes $\ell_s \ll R_0 \ll M_0$ 
for which the correction factor,
\begin{equation}
c \, (R_0) \simeq (1 - c_1 \, R_0^{-2}) \, (1 - c_2 \, R_0^2 / 
M_0^2) \,,
\label{eq4.1}
\end{equation}
in the entropy
\begin{equation}
S (M_0 , R_0) = c \, (R_0) \, a_0 \, M_0 \, , \label{eq4.2}
\end{equation}
is near unity. [We use Eq.~(\ref{eq2.30}) in the limit $g 
\rightarrow 0$, for which it was derived.] Because of this reduced 
sensitivity of $c \,(R_0)$ on a possible direct effect of $g$ on $R$ 
(i.e. $R(g) = R_0 + \delta_g \, R$), the main effect of self-gravity 
on the entropy (considered as a function of the actual values $M$, 
$R$ when $g \not= 0$) will come from replacing $M_0$ as a function 
of $M$ and $R$. The mass-shift result (\ref{eq3.25}) gives $\delta 
\, M = M - M_0$ to first order in $g^2$. To the same 
accuracy\footnote{Actually, Eq.~(\ref{eq4.3}) is probably a more 
accurate version of the mass-shift formula because it
exhibits the real mass $M$ (rather than the bare mass $M_0$) as the
source of self-gravity.}, (\ref{eq3.25}) gives $M_0$ as a function 
of $M$ and $R$:
\begin{equation}
M_0 \simeq M + c_3 \, g^2 \, \frac{M^2}{R^{d-2}} = M \left( 1 + c_3 
\,\frac{g^2 \, M}{R^{d-2}} \right) \, , \label{eq4.3}
\end{equation}
where $c_3$ is a positive numerical constant.

Finally, combining Eqs.~(\ref{eq4.1})--(\ref{eq4.3}) (and 
neglecting, as just said, a small effect linked to $\delta_g \, R 
\not= 0$) leads to the following relation between the entropy, the 
mass and the size (all considered for self-gravitating states, with 
$g \not= 0$)
\begin{equation}
S(M,R) \simeq a_0 \, M \left( 1 - \frac{1}{R^2} \right) \left( 1 -
\frac{R^2}{M^2} \right) \left( 1 + \frac{g^2 \, M}{R^{d-2}} \right) 
\, . \label{eq4.4}
\end{equation}
For notational simplicity, we henceforth set to unity (by using suitable 
redefinitions) the 
coefficients $c_1$, $c_2$ and $c_3$. The possibility of smoothly 
transforming self-gravitating string states into black hole states come from 
the peculiar radius dependence of the entropy $S(M,R)$. Eq.~(\ref{eq4.3}) 
exhibits two effects 
varying in opposite directions: (i) self-gravity favors small
values of $R$ (because they correspond to larger values of $M_0$, 
i.e. of the ``bare'' entropy), and (ii) the constraint of being of 
some fixed size $R$ disfavors both small $(R \ll \sqrt{M})$ and 
large $(R \gg \sqrt{M})$ values of $R$. For given values of $M$ and 
$g$, the most numerous (and therefore most probable) string states 
will have a size $R_* (M;g)$ which maximizes the entropy $S(M,R)$. 
Said differently, the total degeneracy of the complete ensemble of 
self-gravitating string states with total energy $M$ (and no {\it a 
priori} size restriction) will be given by an integral (where 
$\Delta \, R$ is the rms fluctuation of $R$)
\begin{equation}
{\cal D} (M) \sim \int \frac{d R}{\Delta \, R} \, e^{S(M,R)} \sim 
e^{S(M,R_*)} \label{eq4.5}
\end{equation}
which will be dominated by the saddle point $R_*$ which maximizes 
the exponent.

The value of the most probable size $R_*$ is a function of $M$, $g$ 
and the space dimension $d$. We refer to Ref.~\cite{dv} for a full 
treatment. Let us only indicate the results in the (actual) case where 
$d=3$. When maximizing the entropy $S(M,R)$ with respect to $R$ one finds 
that: (i) when $g^2 \ll M^{-3/2}$, the most probable size $R_* (M,g) \sim 
\sqrt{M}$, (ii) when $g^2 \gg M^{-3/2}$,
\begin{equation}
R_* (M,g) \simeq \frac{1 + \sqrt{1+3 \lambda^2}}{\lambda} \label{eq4.1new}
\end{equation}
where $\lambda \equiv g^2 M$.

Eq.~(\ref{eq4.1new}) says that, when $g^2$ increases, and therefore when 
$\lambda$ increases (beyond $M^{-1/2}$) the typical size of a 
self-gravitating string decreases, and (formally) tends to a limiting size 
of order unity, $R_{\infty} = \sqrt 3$ (i.e. of order the string length 
scale $\ell_s = \sqrt{2 \alpha'}$) when $\lambda \gg 1$. However, the 
fractional self-gravity $G_N M / R_* \simeq \lambda / R_*$ (which measures 
the gravitational deformation away from flat space) becomes unity for 
$\lambda = \sqrt 5$ and formally increases without limit when $\lambda$ 
further increases. Therefore, we expect that for some value of $\lambda$ of 
order unity, the self-gravity of the compact string state already reached 
when $\lambda \sim 1$ (indeed, Eq.~(\ref{eq4.1new}) predicts $R_* \sim 1$ 
when $\lambda \sim 1$) will become so strong that it will (continuously) 
turn into a black hole state. Having argued that the dynamical threshold for 
the transition string $\rightarrow$ black hole is $\lambda \sim 1$, we now 
notice that, for such a value of $\lambda$ the entropy $S(M) = S (M,R_* (M)) 
\simeq a_0 \, M \left[ 1 + \frac{1}{4} \, (g^2 M)^2 \right]$ of the string 
state (of mass $M$) matches the Bekenstein-Hawking entropy $S_{\rm BH} (M) 
\sim g^2 \, M^2 = \lambda M$ of the formed black hole. One further checks 
that the other global physical characteristics of the string state (radius 
$R_*$, luminosity $P$, Eq.~(\ref{eq3.29})) match those of a Schwarzschild 
black hole of the same mass ($R_{\rm BH} \sim GM \sim g^2 M \, , \, 
P_{\rm Hawking} \sim R_{\rm BH}^{-2}$) when $\lambda \sim 1$.

\section{Discussion}

Conceptually, the main new result of this paper concerns the most
probable state of a very massive single\footnote{We consider states 
of a single string because, for large values of the mass, the 
single-string entropy approximates the total entropy up to 
subleading terms.} self-gravitating string.  By combining our 
estimates of the entropy reduction due to the size constraint, and 
of the mass shift we come up with the expression (\ref{eq4.4}) for 
the logarithm of the number of self-gravitating 
string states of size $R$.  Our analysis of the function $S(M,R)$ 
clarifies the correspondence~\cite{susskind},~\cite{halyo}, 
\cite{hp1},~\cite{hp2} between string states and black holes.  In 
particular, our results confirm many of the results of~\cite{hp2}, 
but make them (in our opinion) physically clearer by dealing 
directly with the size distribution, in real space, of an ensemble 
of string states.  When our results differ from those of~\cite{hp2}, 
they do so in a way which simplifies the physical picture
and make even more compelling the existence of a correspondence 
between strings and black holes. The simple physical picture 
suggested\footnote{Our conclusions are not rigourously established 
because they rely on assuming the validity of the result 
(\ref{eq4.4}) beyond the domain ($R^{-2} \ll 1$, $g^2 \, M / R^{d-2}
\ll 1$) where it was derived. However, we find heuristically 
convincing to believe in the presence of a reduction factor of the 
type $1-R^{-2}$ down to sizes very near the string scale. Our 
heuristic dealing with self-gravity is less compelling because we do 
not have a clear signal of when strong gravitational field effects 
become essential.} by our results is the following: In any 
dimension, if we start with a massive string state and increase the 
string coupling $g$, a {\it typical} string state will,
eventually, become more compact and will end up, when $\lambda_c = 
g_c^2\, M \sim 1$, in a ``condensed state'' of size $R \sim 1$, and 
mass density $\rho \sim g_c^{-2}$. Note that the basic reason why 
small strings, $R \sim 1$, dominate in any dimension the entropy when 
$\lambda \sim 1$ is that they descend from string states with bare 
mass $M_0 \simeq M (1 + \lambda / R^{d-2}) \sim 2 M$ which are 
exponentially more numerous than less condensed string states 
corresponding to smaller bare masses.

The nature of the transition between the initial ``dilute'' state 
and the final ``condensed'' one depends on the value of
the space dimension $d$. In $d=3$, the transition 
is gradual: when
$\lambda < M^{-1/2}$ the size of a typical state is $R_*^{(d=3)} 
\simeq M^{1/2}(1-M^{1/2} \, \lambda / 8)$, when $\lambda >  M^{1/2}$ 
the typical size is $R_*^{(d=3)} \simeq (1 + (1+3 \, 
\lambda^2)^{1/2}) / \lambda$. In $d=4$, the transition toward a 
condensed state is still continuous, but most of the size evolution 
takes place very near $\lambda = 1$: when $\lambda <1$,
$R_*^{(d=4)} \simeq M^{1/2} (1-\lambda)^{1/4}$, and when $\lambda > 
1$,$R_*^{(d=4)} \simeq (2\lambda / (\lambda - 1))^{1/2}$, with some 
smooth blending between the two evolutions around $\vert \lambda - 1 
\vert \sim M^{-2/3}$. In $d \geq 5$, the transition is discontinuous 
(like a first order phase transition between, say, gas and liquid 
states). Barring the consideration of metastable (supercooled) 
states, on expects that when $\lambda = \lambda_2 \simeq \nu^{\nu} / 
(\nu - 1)^{\nu - 1}$ (with $\nu = (d-2) /2$), the most probable size 
of a string state will jump from $R_{\rm rw}$ (when $\lambda < 
\lambda_2$) to a size of order unity (when $\lambda > \lambda_2$).

One can think of the ``condensed'' state of (single) string matter, 
reached (in any $d$) when $\lambda \sim 1$, as an analog of a 
neutron star with respect to an ordinary star (or a white dwarf). It 
is very compact (because of self gravity) but it is stable (in
some range for $g$) under gravitational collapse. However, if one 
further increases $g$ or $M$ (in fact, $\lambda = g^2 \, M$), the 
condensed string state is expected (when $\lambda$ reaches some 
$\lambda_3 > \lambda_2$, $\lambda_3 = {\cal O} (1)$) to collapse 
down to a black hole state (analogously to a neutron star collapsing 
to a black hole when its mass exceeds the Landau-Oppenheimer-Volkoff 
critical mass). Still in analogy with neutron stars, one notes that 
general relativistic strong gravitational field effects are crucial 
for determining the onset of gravitational collapse; indeed, under 
the ``Newtonian'' approximation (\ref{eq4.4}), the condensed string 
state could continue to exist for arbitrary large values of 
$\lambda$.

It is interesting to note that the value of the mass density at the
formation of the condensed string state is $\rho \sim g^{-2}$. This
is reminiscent of the prediction by Atick and Witten~\cite{AW88} of 
a first-order phase transition of a self-gravitating thermal gas of 
strings, near the Hagedorn temperature\footnote{Note that, by 
definition, in our {\it single} string system, the formal 
temperature $T = (\partial S / \partial M)^{-1}$ is always near the 
Hagedorn temperature.}, towards a dense state with energy
density $\rho \sim g^{-2}$ (typical of a genus-zero contribution to 
the free energy). Ref.~\cite{AW88} suggested that this transition is 
first-order because of the coupling to the dilaton. This suggestion 
agrees with our finding of a discontinuous transition to the single 
string condensed state in dimensions $\geq 5$ (Ref.~\cite{AW88} work 
in higher dimensions, $d=25$ for the bosonic case). It would be 
interesting to deepen these links between self-gravitating single 
string states and multi-string states.

Let us come back to the consequences of the picture brought 
by the present work for the problem of the end point of the 
evaporation of a Schwarzschild black hole and the interpretation of 
black hole entropy. In that case one fixes the value of $g$ (assumed 
to be $\ll 1$) and considers a black hole which slowly looses its 
mass via Hawking radiation. When the mass gets as low as a 
value\footnote{Note that the mass at the black hole $\rightarrow$ 
string transition is larger than the Planck mass $M_P \sim 
(G_N)^{-1/2} \sim g^{-1}$ by a factor $g^{-1} \gg 1$.} $M \sim 
g^{-2}$, for which the radius of the black hole is of order one 
(in string units), one expects the black hole to transform (in all
dimensions) into a typical string\footnote{A check on the 
single-string dominance of the transition black hole $\rightarrow$ 
string is to note that the single string entropy $\sim M / M_s$ is 
much larger than the entropy of a ball of radiation $S_{\rm rad} 
\sim (RM)^{d/(d+1)}$ with size $R \sim R_{\rm BH} \sim \ell_s$ at 
the transition.} state corresponding to $\lambda = g^2 \, M \sim 1$, 
which is a dense state (still of radius $R \sim 1$). This 
string state will further decay and loose mass, predominantly via 
the emission of massless quanta, with a quasi thermal spectrum with 
temperature $T \sim T_{\rm Hagedorn} = a_0^{-1}$ which smoothly matches the 
previous black hole Hawking 
temperature. This mass loss will further decrease the product 
$\lambda = g^2 \, M$, and this decrease will, either gradually or 
suddenly, cause the initially compact string state to inflate to 
much larger sizes. For instance, if $d \geq 4$, the string state
will quickly inflate to a size $R \sim \sqrt{M}$. Later, with 
continued mass loss, the string size will slowly shrink again toward 
$R \sim 1$ until a remaining string of mass $M \sim 1$ finally 
decays into stable massless quanta. In this picture, the black hole 
entropy acquires a somewhat clear statistical significance (as the 
degeneracy of a corresponding typical string state) {\it only} when 
$M$ and $g$ are related by $g^2 \, M \sim 1$. If we allow ourselves 
to vary (in a Gedanken experiment) the value of $g$ this gives a 
potential statistical significance to any black hole entropy value 
$S_{\rm BH}$ (by choosing $g^2 \sim S_{\rm BH}^{-1}$). We do not 
claim, however, to have a clear idea of the direct statistical meaning 
of $S_{\rm BH}$ when $g^2 \, S_{\rm BH} \gg 1$. Neither do we 
clearly understand the fate of the very large space (which could be 
excited in many ways) which resides inside very large classical 
black holes of radius $R_{\rm BH}\sim (g^2 \, S_{\rm BH})^{1/(d-1)} 
\gg 1$. The fact that the interior of a black hole of given mass could 
be 
{\it arbitrarily} large\footnote{E.g., in the Oppenheimer-Snyder 
model, one can join an arbitrarily large closed Friedmann dust universe, 
with hyperspherical opening angle $0 \leq \chi_0 \leq \pi$ arbitrarily 
near 
$\pi$, onto an exterior Schwarzschild spacetime of given mass $M$.}, 
and therefore arbitrarily complex, suggests that black hole physics 
is not exhausted by the idea (confirmed in the present paper) of a 
reversible transition between string-length-size black holes and 
string states.

On the string side, we also do not clearly understand how one could 
follow in detail (in the present non BPS framework) the 
``transformation'' of a strongly self-gravitating string state into 
a black hole state.

Finally, let us note that we expect that self-gravity will lift 
nearly completely the degeneracy of string states. [The degeneracy 
linked to the rotational symmetry, i.e. $2J + 1$ in $d=3$, is 
probably the only one to remain, and it is negligible compared to 
the string entropy.] Therefore we expect that the separation $\delta 
\, E$ between subsequent (string and black hole) energy levels will 
be exponentially small: $\delta \, E \sim \Delta \, M \, \exp 
(-S(M))$, where $\Delta \, M$ is the canonical-ensemble fluctuation 
in $M$. Such a $\delta \, E$ is negligibly small compared to the 
radiative width $\Gamma \sim g^2 \, M$ of the levels. This seems to 
mean that the discreteness of the quantum levels of strongly 
self-gravitating strings and black holes is very much blurred, and 
difficult to see observationally.

\end{document}